\RequirePackage{fix-cm}
\documentclass[smallextended]{svjour3}       
\smartqed  
\usepackage{graphicx}
\begin{document}

\title{ Lyapunov Exponent and Charged Myers Perry Spacetimes 
}


\author{Partha Pratim Pradhan}


\institute{\at Department of Physics\\
           Vivekananda Satabarshiki Mahavidyalaya \\
           Manikpara, Paschim Medinipur\\
            WestBengal-721513, India. \\
            \email{pppradhan77@gmail.com}
}

\date{Received: date / Accepted: date}

\maketitle

\begin{abstract}
We compute the proper time Lyapunov exponent for charged Myers Perry black hole spacetime and investigate 
the instability of the equatorial circular geodesics (both timelike and null) via this exponent. We also 
show that for more than four spacetime dimensions $(N \geq 3)$, there are \emph{no} Innermost Stable Circular 
Orbits (ISCOs) in charged Myers Perry black hole spacetime. We further show that among all possible circular 
orbits, timelike circular orbits have  \emph{longer} orbital periods than null circular orbits (photon spheres) 
as measured by  asymptotic observers. Thus, timelike circular orbits provide the \emph{slowest way} to orbit 
around the charged Myers Perry black hole.


\end{abstract}

\section{Introduction}

Geodesics, especially equatorial circular geodesics in four dimensional $(3+1)$ spacetimes \cite{sch},  have been
extensively discussed in the literature.   In $(N+1)$  spacetime dimensions, such studies are confined to specific goemetries 
inspired to some extent by string theoretic solutions. It has been shown that \cite{mp} higher dimensional black hole spacetimes
have certain  distinct features beyond their four dimensional counterparts. For example, higher dimensional spacetimes admit
black-ring \cite{emparan}, black-string \cite{gl} and black-Saturn \cite{ef} solutions, four dimensional analogues of which 
do not exist.  Again higher dimensional spacetimes  permit a variety of horizon topologies, whereas in $(3+1)$ spacetime the
horizon topologies are usually ${\bf R} \times S^2$.

On the other hand, four dimensional black holes have a number of remarkable features, such as their obeisance of uniqueness 
theorems, dynamical stability, and the laws of black hole mechanics. It is thus important
to investigate what happens  to these issues in higher dimensional spacetimes. Although the laws of black hole mechanics are
known to hold for Lorentzian spacetimes of arbitrary dimension, the uniqueness theorem  is violated in higher dimensional 
spacetimes. In \cite{gt}, the authors argue that there may be a possibility of creation of
higher dimensional black holes in a future Large Hadron Collider, thereby  bringing such black holes into the realm of reality.

Tangerlini \cite{tangher} was the first to derive static black hole solutions of the Einstein equation in $> 4$ dimesions. These solutions
generalize spherically symmetric Schwarzschild and Reissner-Nordstr{\o}m black holes of $3+1$ dimension in Einstein's
general relativity. For higher dimensional static spherically symmetric  black holes, a uniqueness theorem still exists
\cite{gis}.  However, as mentioned above, Myers-Perry black holes present a different situation.

In this paper, we study properties of causal geodesics in charged Myers Perry blackhole spacetime. We further compute the 
proper time Lyapunov exponent for such geodesics in this spacetime. Using  this exponent we shall prove that for
$(N \geq 3)$, there are no ISCOs in charged Myers Perry blackhole space-times. Note  that the principal Lyapunov
exponents($\lambda$) have been computed in \cite{clv,cl,car} using a \emph{coordinate}
time $t$,  where $t$ is measured by the asymptotic observers . Thus, these exponents are explicitly coordinate dependent
and therefore have a degree of unphysicality. Here we compute the principal Lyapunov exponent ($\lambda$)
analytically by using the \emph{proper time} which is coordinate invariant.

Thus the proper time Lyapunov exponnet can be derived as in section \ref{def} as well as in our previous work 
for  Reissner-Nordstr{\o}m black holes \cite{pp3} and Kerr-Newman spacetimes \cite{pp5}:
\begin{eqnarray}
\lambda_{proper} & = & \pm \sqrt{\frac{(\dot{r}^{2})''}{2}} ~.\label{potp}
\end{eqnarray}

The  paper is organized as follows: in section \ref{def} we give the basic definition of Lyapunov exponent and also
show that it may be expressed  in terms of the radial effective potential. In section \ref{cerp} we demonstrate that 
the reciprocal of  the critical exponent can be expressed in terms of the effective radial potential.  In
section \ref{ecgmp}, we fully describe the equatorial circular geodesics, both time-like and null,
for charged Myers Perry space-times. In section \ref{letc}  we show that the Lyapunov exponent can be used
to study the instability of time-like circular geodesics. In section \ref{cetc} we compute the
reciprocal of the Critical exponent  explicitly; we conclude with discussions in section \ref{dis}.

\section{\label{def} Proper time Lyapunov exponent and Radial Potential:}

In any classical phase space the Lyapunov exponent \cite{lya} gives a measure of the average rate of expansion and contraction 
of a trajectories(geodesics) surrounding it.
A positive Lyapunov exponent indicate a divergence between two nearby geodesics, i.e. the paths of such a system are
extremely sensitive to changes of the initial conditions. A negative Lyapunov exponent implies a convergence between
two nearby geodesics and the vanishing Lyapunov exponent indicate the existence of marginal stability.

The rate of exponential expansion or contraction in the direction of $y(0)$ on the trajectory passing through $X_{0}$
(trajectory at $t=0$) is given by

\begin{eqnarray}
\lambda_{i} &=& \lim_{t \rightarrow \infty} \left( \frac{1}{t}\right)\ln
\left( \frac{\parallel y(t) \parallel}{ \parallel y(0) \parallel}\right) ~.\label{si}
\end{eqnarray}
where $\parallel \parallel$ denotes a vector norm. The asymptotic quantity $\lambda_{i}$ is called the Lyapunov exponent.

If there exists a set of $n$ Lyapunov exponents associated with an n-dimensional autonomous system and they can be
ordered by size that is
\begin{eqnarray}
\lambda_{1}\geq \lambda_{2} \geq \lambda_{3}\geq, ... ,\geq \lambda_{n} ~.\label{le}
\end{eqnarray}
The set of n-numbers $\lambda_{i}$ is called the Lyapunov Spectrum \cite{naf}.

For $p$-dimensions, a $p$-dimensional Lyapunov exponent $\lambda $
is defined as

\begin{eqnarray}
\lambda^{p} &=& \lim_{t \rightarrow \infty} \left( \frac{1}{t}\right)\ln
\left( \frac{\parallel y_{1}(t)\wedge y_{2}(t)\wedge ...\wedge y_{p}(t)\parallel}
{\parallel y_{1}(0)\wedge y_{2}(0)\wedge ... \wedge y_{p}(0)\parallel}\right) ~.\label{ly}
\end{eqnarray}
where $\wedge$ is an exterior or vector cross product.

To derive the Lyapunov exponent in terms of the radial equation we shall first derive
the 2nd derivative of the square of the radial component of the four velocity in terms of Lyapunov exponent. Now
the Lagrangian for a test particle in the equatorial plane for any stationary  axi-symmetric space-time can be written as
\begin{eqnarray}
\cal L &=& \frac{1}{2}\left[g_{tt}\,{\dot{t}}^2+2g_{t\phi}\dot{t}\dot{\phi}+g_{rr}\,{\dot{r}}^2
+g_{\phi\phi}\,{\dot{\phi}}^2\right] ~.\label{lagg}
\end{eqnarray}
Now we defining the canonical momenta as
\begin{eqnarray}
p_{q}&=& \frac{\delta {\cal L}}{\delta\dot{q}}~.\label{cm}
\end{eqnarray}

Now from the Euler-Lagrange equations of motion
\begin{eqnarray}
\frac{dp_{q}}{d\tau} &=& \frac{\delta {\cal L}}{\delta q}~.\label{el}
\end{eqnarray}
Using it we get the non-linear differential equation in  2-dimensional phase space with phase space variables $X_{i}(t)=(p_{r},~r)$.
\begin{eqnarray}
\frac{dp_{r}}{d\tau} &=& \frac{\delta {\cal L}}{\delta r} ~~\mbox{and}~~
\frac{dr}{d\tau} = \frac{p_{r}}{g_{rr}}~.\label{drdt}
\end{eqnarray}
Now linearizing the equations of motion about circular orbits of constant $r$, we get the infinitesimal evolution matrix as
\begin{eqnarray}
M_{ij}=\left(
       \begin{array}{cc}
        0 & \frac{d}{dr}\left(\frac{\delta {\cal L}}{\delta r}\right)\\
        \frac{1}{g_{rr}} & 0 \\
      \end{array}
      \right)|_{r=r_{0}}  ~.\label{Mij}
\end{eqnarray}

Now for circular orbits of constant $r=r_{0}$  have the characteristic values of the matrix gives the information about 
stability of the orbits. The eigen values of this matrix are the principal Lyapunov exponent. Therefore the eigen values
of the evolution matrix along the circular orbit can be written as
\begin{eqnarray}
\lambda^2=  \frac{1}{g_{rr}}
\frac{d}{dr}\left(\frac{\delta {\cal L}}{\delta r}\right)|_{r=r_{0}}~.\label{eigen}
\end{eqnarray}
Again from Lagrange's equation of motion
\begin{eqnarray}
 \frac{d}{d\lambda}\left(\frac{\delta {\cal L}}{\delta \dot{r}}\right)
 -\frac{\delta {\cal L}}{\delta r}=0~.\label{L}
\end{eqnarray}
Thus the Lyapunov exponent(which is the inverse of the instability time scale associated with the geodesic motions) 
in terms of the square of the radial velocity ($\dot{r}^2$) can be written as

\begin{eqnarray}
\frac{\delta {\cal L}}{\delta r}=\frac{1}{2g_{rr}}
\frac{d}{dr}\left(\dot{r}g_{rr}\right)^{2}~.\label{dLdr}
\end{eqnarray}
Finally the principal Lyapunov exponent can be rewritten as
\begin{eqnarray}
\lambda^2&=& \frac{1}{2} \frac{1}{g_{rr}}
\frac{d}{dr}\left[\frac{1}{g_{rr}}\frac{d}{dr}\left(\dot{r}g_{rr}\right)^{2}\right]
~.\label{pve}
\end{eqnarray}
Again for circular geodesics \cite{sch}
\begin{eqnarray}
\dot{r}^{2}=(\dot{r}^{2})'=0 ~.\label{cir}
\end{eqnarray}
where prime denotes for a derivative with respect to $r$.
Therefore the equation (\ref{pve}) for proper time Lyapunov exponent \cite{pp3}  must be reduces to
\begin{eqnarray}
\lambda & = & \pm \sqrt{\frac{(\dot{r}^{2})''}{2}} ~.\label{pot}
\end{eqnarray}
where we may defined $\dot{r}^2$ as radial potential or effective radial potential. In general the Lyapunov exponent
come in $\pm$ pairs to conserve the volume of phase space. From now we shall take only positive Lyapunov exponent. The
circular orbit is unstable when the $\lambda$ is real, the circular orbit is stable when the $\lambda$ is imaginary 
and the circular orbit is marginally stable or saddle point when $\lambda=0$.

(Note that in reference  \cite{car}, the authors use a different definition of the Lyapunov exponent(Coordinate time), 
$\lambda = \sqrt{\frac{V_{r}''}{2\dot{t}^2}}$ with $V_{r}=\dot{r}^{2}$ ).

\section{\label{cerp} Critical exponent and Radial potential:}

Following Pretorius and Khurana\cite{pk}, we can define  Critical exponent which is
the ratio of Lyapunov time scale $T_{\lambda}$ and Orbital time scale $T_{\Omega}$ may
be written as
\begin{eqnarray}
\gamma = \frac{\Omega}{2\pi\lambda}
=\frac{T_{\lambda}}{T_{\Omega}}=\frac{Lyapunov \, Time scale}{Orbital \, Time scale}~.\label{ce}
\end{eqnarray}
where we have introduced $T_{\lambda}=\frac{1}{\lambda}$ and $T_{\omega}=\frac{2\pi}{\Omega}$,
which is important for black-hole merger in the ring down radiation.
In terms of the square of the proper radial velocity ($\dot{r}^2$), Critical exponent can be
written as
\begin{eqnarray}
\gamma &=& \frac{T_{\lambda}}{T_{\Omega}}=
\frac{1}{2\pi}\sqrt{\frac{2\Omega ^2}{(\dot{r}^{2})''}} ~.\label{cerd}
\end{eqnarray}

Alternatively the reciprocal of critical exponent is proportional to the effective radial potential which is given by
\begin{eqnarray}
\frac{1}{\gamma}&=&\frac{T_{\Omega}}{T_{\lambda}}=
2\pi\sqrt{\frac{(\dot{r}^{2})''}{2\Omega^2}} ~.\label{ceinv}
\end{eqnarray}

\section{\label{ecgmp} Instability of Equatorial Circular Geodesics  of the  Charged Myers-Perry space-time:}

We will start our journey with a charged Myers-Perry black hole of $N+1$ dimension which rotates in a single plane with
only one non-zero angular momentum parameter $a$ and is a solution of the vacuum Einstein equation. Therefore the space-time
metric in terms of Boyer-Lindquist  coordinates is given by \cite{dianyan,aliev,rabin}
\begin{eqnarray}
ds^2 & = &- \left(1-\frac{mr^{4-N}}{\Sigma}+
\frac{q^2\,r^{2(3-N)}}{\,\Sigma} \right) dt^2-\frac{2 a\left(m
r^{4-N}-q^2 \,r^{2(3-N)}\right) \sin^2 \theta} {\Sigma}\,dt \,d\phi \nonumber \\[4mm] &&
+\left(r^2+a^2+\frac{a^2 \left(m r^{4-N}-q^2\,r^{2(3-N)}\right) \sin^2\theta}
{\Sigma}\right)\sin^2{\theta}\,d\phi^2 + {{r^{\,N-2}\, \Sigma
}\over \Delta}\,dr^2 \nonumber \\[4mm] &&
+ \Sigma \, d\theta^2+ r^2\cos^2{\theta} \, d\Omega_{N-3}^2\,\,, \label{metric}
\end{eqnarray}
where,
\begin{eqnarray}
\Delta &=& r^{N-2}(r^2+a^2)-mr^2+q^2r^{4-N},\label{deltaq1}\\
\Sigma &=& r^2+a^2\cos^2\theta,~\label{sigma}
\end{eqnarray}

\begin{eqnarray}
d\Omega^2_{N-3}=d\chi_1^2+\sin^2\chi_1[d\chi^2_2+\sin^2\chi_2(\cdot\cdot\cdot d\chi^2_{N-3})].~\label{omega}
\end{eqnarray}

The electromagnetic potential one form for the space-time (\ref{metric}) is
\begin{eqnarray}
A=A_{\mu}dx^{\mu}=-\frac{Q}{(N-2)r^{N-4}\Sigma}(dt-a\sin^2\theta d\phi).~\label{phi}
\end{eqnarray}

The determinant ($g$) of the metric (\ref{metric}) gives
\begin{eqnarray}
\sqrt{-g}=\sqrt{\gamma}\Sigma r^{N-3}\sin\theta\cos^{N-3}\theta,~\label{deter}
\end{eqnarray}

where $\gamma$ is the determinant of the metric (\ref{omega}).

The parameters $m,~a,~q$ are related with the physical mass ($M$), angular momentum ($J$) and charge ($Q$) through the
relations given by
\begin{eqnarray}
M &=&\frac{A_{N-1}(N-1)}{16\pi G}m .~\label{m}\\
J &=&\frac{A_{N-1}}{8\pi G}ma .~\label{j}\\
Q^2 &=&\frac{(N-2)(N-1)A_{N-1}}{8\pi G}q^2 .~\label{q1}
\end{eqnarray}

Here $A_{N-1}$ is the area of the unit sphere in $N-1$ dimensions
\begin{eqnarray}
A_{N-1}=\int_{0}^{2\pi}d\phi\int_0^\pi \sin\theta\cos^{N-3}\theta d\theta\prod_{i=1}^{N-3}\int_{0}^{\pi}\sin^{(N-3)-i}\chi_{i}d\chi_{i}=
\frac{2\pi^{N/2}}{\Gamma(N/2)}. ~\label{a}
\end{eqnarray}

The position of the event horizon is represented by the largest root of the polynomial $\Delta_{r=r_+}=0$. The angular 
velocity at the event horizon is given by
\begin{eqnarray}
\Omega_{H}=\frac{a}{r_+^2+a^2} .~\label{ang1}
\end{eqnarray}

The semiclassical Bekenstein-Hawking entropy reads\cite{rabin}
\begin{eqnarray}
S=\frac{A_H}{4}=\frac{\pi r_+^{N-3}(r_+^2+a^2)\bar{A}_{N-3}}{N-2}.
\label{entp}
\end{eqnarray}
where
\begin{eqnarray}
\bar{A}_{N-3}=\prod_{i=1}^{N-3}\int_{0}^{\pi}\sin^{(N-3)-i}\chi_{i}d\chi_{i}=\frac{\pi^{\frac{N}{2}-1}}{\Gamma(\frac{N}{2}-1)}.
\label{an-3}
\end{eqnarray}

The surface gravity of this space-times (\ref{metric}) is given by
\begin{eqnarray}
\kappa &=& \frac{(N-4)(r_+^2+a^2)+2r_+^2-(N-2)q^2r_+^{2(3-N)}}{2 r_+(r_+^2+a^2)}.
\label{sg}
\end{eqnarray}

and  the Hawking temperature reads
\begin{eqnarray}
T_{H} &=& \frac{\kappa}{2\pi} = \frac{(N-4)(r_+^2+a^2)+2r_+^2-(N-2)q^2r_+^{2(3-N)}}{4\pi r_+(r_+^2+a^2)}.
\label{hawt}
\end{eqnarray}

In appropriate limits the metric (\ref{metric}), the BH entropy (\ref{entp}), surface gravity (\ref{sg}) and  Hawking temperature
(\ref{hawt}) reproduces the $N+1$ dimensional spherically symmetric, static Schwarzscild,
Reissner-Nordstr{\o}m \cite{tangher} and axially symmetric,  Myers-Perry spcetime\cite{mp}.

To compute the geodesics in the equatorial plane for the Charged Myers Perry  space-time we follow\cite{sch}. To determine the 
geodesic motions of a test particle in this plane we set $\dot{\theta}=0$ and $\theta=constant=\frac{\pi}{2}$.

Therefore the necessary Lagrangian for this motion is given by
\begin{eqnarray}
2{\cal L} &=&
 -\left(1-\frac{m}{r^{N-2}}+\frac{q^2}{r^{2N-4}}\right)\,\dot{t}^{2} -\left(\frac{2ma}{r^{N-2}}-\frac{2aq^2}{r^{2N-4}}\right)\dot{t}\,\dot{\phi}+
 \frac{r^N}{\Delta}\dot{r}^2+ \nonumber \\[4mm] &&
 \left(r^2+a^2+\frac{ma^2}{r^{N-2}}-\frac{a^2q^2}{r^{2N-4}}\right)\dot{\phi}^2  ~.\label{hamh1}
\end{eqnarray}

The generalized momenta can be derived from it are
\begin{eqnarray}
p_{t} &=& -\left(1-\frac{m}{r^{N-2}}+\frac{q^2}{r^{2N-4}}\right)\,\dot{t} -\left(\frac{ma}{r^{N-2}}-\frac{aq^2}{r^{2N-4}}\right)\dot{\phi} =-E =Const ~.\label{pt}\\
p_{\phi} &=& -\left(\frac{ma}{r^{N-2}}-\frac{aq^2}{r^{2N-4}}\right)\,\dot{t}+  \left(r^2+a^2+\frac{ma^2}{r^{N-2}}-\frac{a^2q^2}{r^{2N-4}}\right)\,\dot{\phi}=L=Const ~.\label{pphi}\\
p_{r} &=& \frac{r^N}{\Delta} \, \dot{r}  ~.\label{pr}
\end{eqnarray}
Here $(\dot{t},~\dot{r},~\dot{\phi})$ denotes differentiation with respect to proper time($\tau$). Since the Lagrangian does not
depends on `t' and `$\phi$', so $p_{t}$ and $p_{\phi}$ are conserved quantities. The independence of the Lagrangian on `t' and
`$\phi$' manifested, the stationarity and the axisymmetric character of the Charged Myers Perry space-time.
The Hamiltonian is given by
\begin{eqnarray}
\cal H &=& p_{t}\,\dot{t}+p_{\phi}\,\dot{\phi}+p_{r}\,\dot{r}-\cal L ~.\label{hamil}
\end{eqnarray}

In terms of the metric the Hamiltonian is
\begin{eqnarray}
2\cal H &=& -\left(1-\frac{m}{r^{N-2}}+\frac{q^2}{r^{2N-4}}\right)\,\dot{t}^{2} -\left(\frac{2ma}{r^{N-2}}-\frac{2aq^2}{r^{2N-4}}\right)\dot{t}\,\dot{\phi}+
 \nonumber \\[4mm] &&
\frac{r^N}{\Delta}\dot{r}^2+
\left[r^2+a^2+\frac{ma^2}{r^{N-2}}-\frac{a^2q^2}{r^{2N-4}}\right]\dot{\phi}^2  ~.\label{hamh}
\end{eqnarray}
Since the Hamiltonian is independent of `t', therefore we can write it as

\begin{eqnarray}
2\cal H &=& -\left[\left( 1-\frac{m}{r^{N-2}}+\frac{q^2}{r^{2N-4}}\right)\,\dot{t}+
\left( \frac{ma}{r^{N-2}}-\frac{aq^2}{r^{2N-4}}\right)\,\dot{\phi} \right]\,\dot{t}+\frac{r^N}{\Delta}\,\dot{r}^2
  +\nonumber \\[4mm] && \left[\left(\frac{ma}{r^{N-2}}-\frac{aq^2}{r^{2N-4}} \right)\,\dot{t}+\left(r^2+a^2+\frac{ma^2}{r^{N-2}}-\frac{a^2q^2}
  {r^{2N-4}}\right)\,\dot{\phi} \right]\,\dot{\phi} ~.\label{2ham}\\ &=&-E\,\dot{t}+L\,\dot{\phi}+\frac{r^N}{\Delta}\,\dot{r}^2=\epsilon=const ~.\label{2hd}
\end{eqnarray}
Here $\epsilon=-1$ for time-like geodesics, $\epsilon=0$ for light-like geodesics and $\epsilon=+1$ for space-like geodesics.
Solving equations (\ref{pt}) and (\ref{pphi}) for $\dot{\phi}$ and $\dot{t}$, we find
\begin{eqnarray}
\dot{\phi}&=&\frac{r^{N-2}}{\Delta}\left[\left(1-\frac{m}{r^{N-2}}+\frac{q^2}{r^{2N-4}}  \right)L+\left(\frac{ma}{r^{N-2}}-\frac{aq^2}{r^{2N-4}}\right)E\right] ~.\label{uphi}\\
\dot{t}&=&\frac{r^{N-2}}{\Delta}\left[\left( r^2+a^2+\frac{ma^2}{r^{N-2}}-\frac{a^2q^2}
{r^{2N-4}}\right)E-\left(\frac{ma}{r^{N-2}}-\frac{aq^2}{r^{2N-4}}\right)L\right] ~.\label{ut}
\end{eqnarray}

Inserting these solutions in equations (\ref{2hd}), we obtain the radial equation for Charged Myers Perry space-time 
 is given by
\begin{eqnarray}
r^2\dot{r}^{2} &=& r^2E^2+\left(\frac{m}{r^{N-2}}-\frac{q^2}{r^{2N-4}}\right)\left(aE-L\right)^2 +\left(a^2E^2-L^2\right)+\epsilon \frac{\Delta}{r^{N-2}} ~.\label{radial}
\end{eqnarray}

\subsection{Circular Null Geodesics:}

For null geodesics $\epsilon=0$, introducing new quantities $m=2M$ and $q=Q$ for simplicity the radial equation (\ref{radial}) becomes

\begin{eqnarray}
r^2\dot{r}^{2} &=& r^2E^2+\left(\frac{2M}{r^{N-2}}-\frac{Q^2}{r^{2N-4}} \right)\left(aE-L\right)^2 +\left(a^2E^2-L^2\right) ~.\label{nul}
\end{eqnarray}
The equations finding the radius of $r_{c}$ of the unstable circular `photon orbit' at $E=E_{c}$ and $L=L_{c}$  are
\begin{eqnarray}
E_{c}^2r_{c}^{2}+\left(\frac{2M}{r_{c}^{N-2}}-\frac{Q^2}{r_{c}^{2N-4}}\right)\left(aE_{c}-L_{c}\right)^2
+\left(a^2E_{c}^2-L^2\right)&=& 0 ~.\label{rc}\\
2r_{c}E_{c}^2+\left(-(N-2)\frac{2M}{r_{c}^{N-1}}+(2N-4)\frac{Q^2}{r_{c}^{2N-3}}\right)
\left(aE_{c}-L_{c}\right)^2 &=& 0 ~.\label{rc1}
\end{eqnarray}
Now introducing the impact parameter $D_{c}=\frac{L_{c}}{E_{c}}$, the above equations may be
written as
\begin{eqnarray}
r_{c}^{2}+\left(\frac{2M}{r_{c}^{N-2}}-\frac{Q^2}{r_{c}^{2N-4}}\right)\left(a-D_{c}\right)^2
+\left(a^2-D_{c}^2\right)&=& 0  ~.\label{dc}\\
r_{c}-\left((N-2)\frac{M}{r_{c}^{N-1}}-(N-2)\frac{Q^2}{r_{c}^{2N-3}}\right)\left(a-D_{c}
\right)^2 &=& 0 ~.\label{dc1}
\end{eqnarray}
From equation (\ref{dc1}) we have
\begin{eqnarray}
D_{c}&=& a\mp\frac{r_{c}^{N-1}}{\sqrt{M(N-2)r_{c}^{N-2}-(N-2)Q^2}} ~.\label{dc2}
\end{eqnarray}
The equation (\ref{dc}) is valid if and only if $\mid D_{c}\mid>a$. For counter rotating
orbit, we have $|D_{c}-a|=-(D_{c}-a)$, which correspond to upper sign in the above equation and co-rotating
$|D_{c}-a|=+(D_{c}-a)$,which correspond to lower sign in the above equation. Inserting equation (\ref{dc2})
in (\ref{dc}) we find an equation for the \emph{radius of null circular orbits}
\begin{eqnarray}
r_{c}^{2N-4}-N Mr_{c}^{N-2}\pm 2ar_{c}^{N-3}\sqrt{(N-2)Mr_{c}^{N-2}-(N-2)Q^2}+(N-1)Q^2 &=& 0 ~.\nonumber\\
\label{rnul}
\end{eqnarray}
When $N=3$, we recover the well known photon sphere \cite{cve} equation for the Kerr Newman space-times\cite{pp5}.
Another important relation can be derived using equations (\ref{dc}) and (\ref{dc2}) for null circular orbits are
\begin{eqnarray}
D_{c}^2&=& a^2+r_{c}^2\left(\frac{N Mr_{c}^{N-2}-(N-1)Q^2}{(N-2)Mr_{c}^{N-2}-(N-2)Q^2}\right) ~.\label{dc3}
\end{eqnarray}

Now we will derive an important quantity associated with the circular null geodesics
is the angular frequency which is denoted by $\Omega_{c}$
\begin{eqnarray}
\Omega_{c}=\frac{\left[\left(1-\frac{2M}{r_{c}^{N-2}}+\frac{Q^2}{r_{c}^{2N-4}}\right)D_{c}
+\left(\frac{2M}{r_{c}^{N-2}}-\frac{Q^2}{r_{c}^{2N-4}}\right)a\right]}{\left[\left(r_{c}^2+
a^2+\frac{2Ma^2}{r_{c}^{N-2}}-\frac{a^2Q^2}{r_{c}^{2N-4}}\right)
-a\left(\frac{2M}{r_{c}^{N-2}}-\frac{Q^2}{r_{c}^{2N-4}}\right)D_{c}\right]}=\frac{1}{D_{c}}
~.\label{omnul}
\end{eqnarray}
Using equations  (\ref{dc2}) and (\ref{dc}) we show that the angular frequency $\Omega_{c}$ of the circular null 
geodesics is inverse of the impact parameter $D_{c}$, which generalizes the result of Kerr Newmann case\cite{pp5} to
the charged Myers Perry black-hole space-time. It proves that this is a \emph{general feature} of any  higher 
dimensional stationary space-time.

\subsection{Circular Timelike Geodesics:}

The  time-like geodesics equation (\ref{radial}) can be written as by setting $\epsilon=-1$
\begin{eqnarray}
r^2\dot{r}^2 &=& r^2E^2+\left(\frac{2M}{r^{N-2}}-\frac{Q^2}{r^{2N-4}}\right)\left(aE-L\right)^2
+\left(a^2E^2-L^2\right)-\frac{\Delta}{r^{N-2}} ~.\label{rtime}
\end{eqnarray}
where $E$ is the energy per unit mass of the particle describes the trajectory.

Now we shall find the radial equation of the timelike circular geodesics in terms of reciprocal radius $u=1/r$ as
the independent variable, can be expressed as
\begin{eqnarray}
{\cal V}(u) &=& u^{-4}\dot{u}^2=E^2+2Mu^N\left(aE-L\right)^2-u^{2N-2}Q^2\left(aE-L\right)^2 \nonumber \\[2mm] &&
+\left(a^2E^2-L^2\right)u^2-1-a^2u^2+2Mu^{N-2}-Q^2u^{2N-4} ~.\label{vu}
\end{eqnarray}

The conditions for the occurrence of circular orbits at $r=r_{0}$ or reciprocal radius
$u=u_{0}$ are

\begin{eqnarray}
{\cal V}(u) &=& 0 ~.\label{vu0}
\end{eqnarray}
and
\begin{eqnarray}
\frac{d{\cal V}(u)}{du} &=& 0 ~.\label{dvdu}
\end{eqnarray}
Now setting $x=L_{0}-aE_{0}$, where $L_{0}$ and $E_{0}$ are the values of energy and angular
momentum for circular orbits at the radius $r_{0}=\frac{1}{u_{0}}$.
Therefore using (\ref{vu},~\ref{dvdu}) we get the following equations
\begin{eqnarray}
-x^2Q^2u^{2N-2}+2Mx^2u_{0}^N-(x^2+2axE)u^2-a^2u_{0}^2-Q^2u_{0}^{2N-4}
\nonumber \\[2mm]
+2Mu_{0}^{N-2}-1+E_{0}^2 &=& 0  ~.\label{x1}
\end{eqnarray}
and

\begin{eqnarray}
-(N-1)x^2Q^2u_{0}^{2N-3}+NMx^2u_{0}^{N-1}-(x^2+2axE_{0})u_{0}-a^2u_{0}
\nonumber \\[2mm]
-(N-2)Q^2u_{0}^{2N-5}+(N-2)Mu_{0}^{N-3} &=& 0  ~.\label{x2}
\end{eqnarray}
Using (\ref{x1},~\ref{x2}) we find an equation for $E_{0}^2$ as
\begin{eqnarray}
E_{0}^2 &=& 1+(N-4)Mu_{0}^{N-2}+(N-2)Mx^2u_{0}^N \nonumber\\[2mm]&&
-(N-2)x^2Q^2u_{0}^{2N-2}-(N-3)Q^2u_{0}^{2N-4} ~.\label{e2}
\end{eqnarray}
with the aid of equation (\ref{e2}), equation (\ref{x2}) gives us

\begin{eqnarray}
2axE_{0}u_{0} &=& x^2[NMu_{0}^{N-1}-(N-1)Q^2u_{0}^{2N-3}-u_{0}]-a^2u_{0}-(N-2)Q^2u_{0}^{2N-5}
\nonumber \\[2mm] &&
+(N-2)Mu_{0}^{N-3}  ~.\label{axe}
\end{eqnarray}
Eliminating $E_{0}$ between these equations, we have obtained the following
quadratic equation for $x^2$ i.e
\begin{eqnarray}
{\cal A}x^4+{\cal B}x^{2}+{\cal C} &=& 0   ~.\label{quad}
\end{eqnarray}
where
\begin{eqnarray}
{\cal A} &=& u_{0}^2 \left[N Mu_{0}^{N-1}-(N-1)Q^2u_{0}^{2N-3}-u_{0} \right]^2-4a^2u_{0}^{2}\left[(N-2)Mu_{0}^N-(N-2)Q^2u_{0}^{2N-2}\right]  \nonumber\\
{\cal B} &=& -2 \left[a^{2}u_{0}+(N-2)Q^2U_{0}^{2N-5}-(N-2)Mu_{0}^{N-3}\right]\left[NMu_{0}^{N-1}
-(N-1)Q^{2}u_{0}^{2N-3}-u_{0}\right]
\nonumber \\[2mm] &&
-4a^{2}u_{0}^{2}\left[1+(N-4)Mu_{0}^{N-2}-(N-3)Q^{2}u_{0}^{2N-4}\right]
\nonumber\\
{\cal C} &=& \left[ a^2u_{0}+(N-2)Q^2u_{0}^{2N-5}-(N-2)Mu_{0}^{N-3}\right]^2 \nonumber
\end{eqnarray}
The solution of this equations (\ref{quad}) are
\begin{eqnarray}
x^2 &=& \frac{-{\cal B} \pm {\cal D}}{2 {\cal A}}   ~.\label{root}
\end{eqnarray}
where the discriminant of this equation is
\begin{eqnarray}
{\cal D} &=& 4a\,\Delta_{u_{0}} \sqrt{(N-2)Mu_{0}^{N}-(N-2)Q^2u_{0}^{2N-2}}    ~.\label{discri}
\end{eqnarray}
and
\begin{eqnarray}
\Delta_{u_{0}} &=& 1+a^2u_{0}^{2}-2Mu_{0}^{N-2}+Q^2u_{0}^{2N-4}~.\label{deltau}
\end{eqnarray}
The solutions becomes simpler form by writing

\begin{eqnarray}
\left[1-N Mu_{0}^{N-2}+(N-1)Q^2u_{0}^{2N-4}\right]^2
\nonumber \\[2mm]
-4a^2\left[(N-2)Mu_{0}^N-(N-2)Q^2u_{0}^{2N-2}\right] &=& Z_{+}\,Z_{-}  ~.\label{z+}
\end{eqnarray}
where
\begin{eqnarray}
Z_{\pm} &=& \left[1-N Mu_{0}^{N-2}+(N-1)Q^2u_{0}^{2N-4}\right] \pm
\nonumber \\[2mm] &&
2a\sqrt{\left[(N-2)Mu_{0}^N-(N-2)Q^2u_{0}^{2N-2}\right]}   ~.\label{z+-}
\end{eqnarray}
Thus we get the solution as
\begin{eqnarray}
x^2u_{0}^2 &=& \frac{-{\cal B} \pm {\cal D}}{Z_{+} Z_{-}}   ~.\label{x2u2}
\end{eqnarray}
Thus we find
\begin{eqnarray}
x^2u_{0}^2 &=& \frac{\Delta_{u_{0}}-Z_{\mp}}{Z_{\mp}}   ~.\label{zz}
\end{eqnarray}
Again we can write
\begin{eqnarray}
\Delta_{u_{0}}-Z_{\mp} &=& \left[au_{0} \pm \sqrt{(N-2)Mu_{0}^{N-2}-(N-2)Q^2u_{0}^{2N-4}} \right]^2  ~.\label{za}
\end{eqnarray}
Therefore the solutions for $x$ thus may be written as
\begin{eqnarray}
x &=& - \frac{\left[a\sqrt{u_{0}} \pm \sqrt{(N-2)Mu_{0}^{N-3}-(N-2)Q^2u_{0}^{2N-5}} \right]}{\sqrt{u_{0}{Z}_{\pm}}}  ~.\label{solx}
\end{eqnarray}
Here the upper sign in the foregoing equations applies to counter-rotating orbits, while the lower sign applies to
co-rotating orbits.
Replacing the solution (\ref{solx}) for $x$ in equation (\ref{e2}), we obtain
the energy

\begin{eqnarray}
E_{0} &=&  \frac{1}{\sqrt{{Z}_{\mp}}}\left[1-2Mu_{0}^{N-2}\mp a\sqrt{(N-2)Mu_{0}^{N}-(N-2)Q^2u_{0}^{2N-2}} +Q^{2}u_{0}^{2N-4}\right]
\nonumber\\ 
~.\label{eng}
\end{eqnarray}
and the value of angular momentum associated with the circular orbit is given
by

\begin{eqnarray}
L_{0}=\mp \frac{1}{\sqrt{u_{0}{Z}_{\mp}}}\left[(1+a^2u_{0}^2)\sqrt{(N-2)Mu_{0}^{N-3}-(N-2)Q^2u_{0}^{2N-5}}
\pm 2aM\sqrt{u_{0}^{2N-3}} \mp aQ^2\sqrt{u_{0}^{4N-7}}\right] \nonumber\\
~.\label{ang}
\end{eqnarray}

As we previously defined $E_{0}$ and $L_{0}$ followed by equations (\ref{eng}) and (\ref{ang}) are the energy and the angular
momentum per unit mass of a particle describing a circular orbit of radius $u$.

To compute the stability of timelike circular orbit we must calculate the 2nd order derivative of effective potential
with respect to u for the values of $E_{0}$ and $L_{0}$ specific to circular orbits.

Now the 2nd order derivative of effective potential becomes

\begin{eqnarray}
\frac{d^{2}{\cal V}}{du^{2}}=(N-2)u^{N-4}\left[\left(
NM-2(N-1)Q^{2}u^{N-2}\right)-2(N-3)Q^2u^{N-2}+(N-4)M\right] ~.\label{der2}
\end{eqnarray}
Using (\ref{zz}) we find
\begin{eqnarray}
\frac{d^{2}{\cal V}}{du^{2}}|_{u=u_{0}} &=& \frac{2(N-2)u_{0}^{N-4}}{{Z}_{\mp}}
\left[ \left(NM-2(N-1)Q^2u_{0}^{N-2}\right)\Delta_{u_{0}}-\left(4M-4Q^2u_{0}^{N-2}\right)Z_{\mp} \right] \nonumber\\
~.\label{derf}
\end{eqnarray}
The 2nd order derivative of effective potential shows that it explicitly depends on space-time dimensionality $N$. So, to determine
the stability of equatorial circular geodesics we must check the sign of 2nd order derivative of the function ${\cal V}(u)$
which will be helpful to distinguish between different values of $N$.
Since $E_{0}$, $L_{0}$ and $x=L_{0}-aE_{0}$ must be real, the function $\Delta_{u}$ and ${Z}_{\mp}$
are such that
\begin{eqnarray}
\Delta_{u_{0}}\geq {Z}_{\pm} &\geq & 0 ~.\label{delu}
\end{eqnarray}

Case I: For $N\geq 4$ i.e five dimensional case, the above equation leads to
\begin{eqnarray}
\left[NM-2(N-1)Q^2u_{0}^{N-2}\right] \Delta_{u_{0}} & \geq & \left[4M-4Q^2u_{0}^{N-2}\right]Z_{\mp} ~.\label{del1}
\end{eqnarray}
which immediately suggests that
\begin{eqnarray}
\frac{d^{2}{\cal V}}{du^{2}}& \geq & 0  ~.\label{de11}
\end{eqnarray}
Thus we conclude that there are no ISCO and stable timelike circular orbit around the rotating five-dimensional
charged Myers-Perry blackhole space-time, at least in the ``equatorial'' planes. Which generalizes the previous work
by Frolov and Stojkovic\cite{fs} on five dimensional rotating black hole.

Case II: Now in four dimension $N=3$, the above equation (\ref{derf}) reduces to
\begin{eqnarray}
\frac{d^{2}{\cal V}}{du^{2}}|_{u=u_{0}} &=& \frac{2}{u_{0}{Z}_{\mp}}\left[\left(3M-4Q^2u_{0}\right) \Delta_{u_{0}}-\left(4M-4Q^2u_{0} \right)Z_{\mp}  \right] ~.\label{der1}
\end{eqnarray}
For $\Delta_{u_{0}}\geq {Z}_{\pm}\geq 0$, the equation (\ref{der1}) leads to
\begin{eqnarray}
\left(3M-4Q^2u_{0}\right) \Delta_{u_{0}} < \left(4M-4Q^2u_{0} \right)Z_{\mp} ~.\label{del2}
\end{eqnarray}
which implies

\begin{eqnarray}
\frac{d^{2}{\cal V}}{du^{2}} &<& 0  ~.\label{de1}
\end{eqnarray}

This suggests  that there exist ISCO  and stable timelike circular orbit around the rotating four dimensional
Kerr-Newman space-time\cite{pp5}.

Case III. For $N \geq 3$ i.e arbitrary dimension, the above equation (\ref{derf}) leads to
\begin{eqnarray}
\left[NM-2(N-1)Q^2u_{0}^{N-2}\right] \Delta_{u_{0}} & \geq & \left[4M-4Q^2u_{0}^{N-2}\right]Z_{\mp} ~.\label{del22}
\end{eqnarray}
which immediately suggests that
\begin{eqnarray}
\frac{d^{2}{\cal V}(u)}{du^{2}}& \geq & 0  ~.\label{de2}
\end{eqnarray}
Thus we conclude that in space-time dimensions greater than four i.e $N\geq3$, there are no ISCOs and stable timelike circular
orbits around the rotating higher dimensional charged Myers-Perry blackhole space-time, at least in the ``equatorial''
planes. Which generalizes the previous work by Cardoso\cite{car} on higher dimensional Myers-Perry blackhole space-time.

This may suggested that the absence of bounded stable circular orbit in the black hole
exterior is a generic property of higher dimensional charged black holes. Which generalizes the previous work by
Tangerlini\cite{tangher} on non-rotating higher dimensional black hole and by Frolov and Stojkovic\cite{fs} on 
five dimensional rotating black hole.

\subsubsection{Angular velocity of Timelike Circular Orbit}

Now we compute the orbital angular velocity for timelike circular geodesics at $r=r_{0}$ is given by
\begin{eqnarray}
\Omega_{0}=\frac{\dot{\phi}}{\dot{t}}
      =\frac{\left[\left(1-\frac{2M}{r_{0}^{N-2}}+\frac{Q^2}{r_{0}^{2N-4}}\right)L_{0}
       +\left(\frac{2M}{r_{0}^{N-2}}-\frac{Q^2}{r_{0}^{2N-4}}\right)aE_{0}\right]}
       {\left[\left(r_{0}^2+a^2+\frac{2Ma^2}{r_{0}^{N-2}}-\frac{a^2Q^2}{r_{0}^{2N-4}}\right)E_{0}
       -a\left(\frac{2M}{r_{0}}-\frac{Q^2}{r_{0}^2}\right)L_{0}\right]}~.\label{omet}
\end{eqnarray}
Again this can be rewritten as
\begin{eqnarray}
\Omega_{0}=\frac{\left[ L_{0}-2Mu_{0}^{N-2}x+Q^2u_{0}^{2N-4}x \right]u_{0}^2}{(1+a^2u_{0}^2)E_{0}-2axu_{0}^2(2Mu_{0}^{N-2}-Q^2u_{0}^{2N-4})}~.\label{omeu}
\end{eqnarray}
Now the previously mentioned expression can be simplified as
\begin{eqnarray}
L_{0}-2Mxu_{0}^{N-2}+Q^2xu_{0}^{2N-4} &=& \mp \frac{\sqrt{(N-2)Mu_{0}^{N-3}-(N-2)Q^2u_{0}^{2N-5}}}{\sqrt{u_{0}{Z_{\mp}}}}\Delta_{u_{0}}~.\label{os}\\
(1+a^2u_{0}^2)E_{0}-2aMxu_{0}^{N}+axQ^2u_{0}^{2N-2} &=&
\frac{\Delta_{u_{0}}}{Z_{\mp}}\left[1 \mp a\sqrt{(N-2)Mu_{0}^{N}-(N-2)Q^2u_{0}^{2N-2}}\right]~.\nonumber\\
\label{os1}
\end{eqnarray}
Substituting (\ref{os}) and  (\ref{os1}) into  (\ref{omeu}) we get the angular velocity for timelike circular  geodesics
is given by
\begin{eqnarray}
\Omega_{0}=\mp \frac{\sqrt{(N-2)Mu_{0}^{N}-(N-2)Q^2u_{0}^{2N-2}}}{1 \mp a\sqrt{(N-2)Mu_{0}^{N}-(N-2)Q^2u_{0}^{2N-2}}}~.\label{omef}
\end{eqnarray}

\subsubsection{Ratio of Angular velocity of time like circular orbit to Null Circular Orbit}

Since we have already proved that for timelike circular geodesics the angular velocity is given by the equation (\ref{omef})

Again we obtained for circular null geodesics $\Omega_{c}=\frac{1}{D_{c}}$, so we can deduce similar expression for it is
given by
\begin{eqnarray}
\Omega_{c}=\mp \frac{\sqrt{(N-2)Mu_{c}^{N}-(N-2)Q^2u_{c}^{2N-2}}}
{1\mp a\sqrt{(N-2)Mu_{c}^{N}-(N-2)Q^2u_{c}^{2N-2}}} ~.\label{omec}
\end{eqnarray}
Resultantly we obtain the ratio of angular frequency for time-like circular  geodesics to the angular frequency for null
circular geodesics is
\begin{eqnarray}
\frac{\Omega_{0}}{\Omega_{c}}&=& \left(\frac{\sqrt{Mr_{0}^{N-2}-Q^2}}{\sqrt{Mr_{c}^{N-2}-Q^2}}\right)
\left(\frac{r_{c}^{N-1}\mp a\sqrt{(N-2)Mr_{c}^{N-2}-(N-2)Q^2}}{r_{0}^{N-1}\mp a\sqrt{(N-2)Mr_{0}^{N-2}-(N-2)Q^2}}\right) ~.\label{ratio}
\end{eqnarray}
which is proportional to the radial coordinates $r_{0}$.
For $r_{0}=r_{c}$, $\Omega_{0}=\Omega_{c}$,i.e , when the radius of time-like circular geodesics is equal to the radius of
null circular geodesics, the angular frequency corresponds to that geodesic are equal, which demands that the intriguing
physical phenomena could occur in the curved four dimensional space-time, for example, possibility of exciting
Quasi Normal Modes(QNM) by orbiting particles, possibly leading to instabilities of the curved space-time\cite{car}.
It would be very interesting to investigate such phenomenon occur in higher dimensional space-time.

For $r_{0}>r_{c}$, we  proved that for Schwarzschild black-hole, Reissner Nordstr{\o}m black-hole \cite{pp3} and Kerr
Black-hole the null circular geodesics have the largest angular frequency as measured by asymptotic observers than the
time-like circular geodesics. We therefore conclude that null circular geodesics provide the fastest way to circle 
black holes\cite{hod}. This generalizes the case of axi-symmetric symmetry Kerr Newmann space-time\cite{pp5} to the 
more general case of stationary, axisymmetry  charged Myers Perry blackhole space-times.

Now the ratio of time period for time-like circular  geodesics to the  time period for
null circular geodesics is given by
\begin{eqnarray}
\frac{T_{0}}{T_{c}}&=& \left(\frac{\sqrt{Mr_{c}^{N-2}-Q^2}}{\sqrt{Mr_{0}^{N-2}-Q^2}}\right)
\left(\frac{r_{0}^{N-1}\mp a\sqrt{(N-2)Mr_{0}-(N-2)Q^2}}{r_{c}^{N-1}\mp a\sqrt{(N-2)Mr_{c}-(N-2)Q^2}}\right) ~.\label{ratio1}
\end{eqnarray}
This ratio is valid for $r_{0}\neq r_{c}$. For $r_{0}=r_{c}$, $T_{0}=T_{c}$, i.e. time period of both geodesics are equal, which 
possibly leading to the excitations of Quasi Normal Modes.
For $r_{0}>r_{c}$, $T_{0}>T_{c}$, which implies that the orbital period for time-like circular geodesics is greater than the
orbital period for null circular geodesics. For $r_{0}=r_{particle}$ and $r_{c}=r_{photon}$, therefore the ratio of time 
period for the orbit of massive particles ($r_{0}=r_{timelike}$) to the  time period for photon-sphere $(r_{c}=r_{photon})$ for
charged Myers Perry  black-hole is given by
\begin{eqnarray}
\frac{T_{particle}}{T_{photon}}&=& \left(\frac{\sqrt{Mr_{c}^{N-2}-Q^2}}{\sqrt{Mr_{0}^{N-2}-Q^2}}\right)
\left(\frac{r_{0}^{N-1}\mp a\sqrt{(N-2)Mr_{0}-(N-2)Q^2}}{r_{c}^{N-1}\mp a\sqrt{(N-2)Mr_{c}-(N-2)Q^2}}\right) ~.\label{isphkn}
\end{eqnarray}
This implies that $T_{particle}>T_{photon}$, therefore we conclude that timelike circular geodesics provide the \emph{slowest way}
to circle the charged Myers Perry black-hole space-time.

Here we may note that we recover from (\ref{rnul}) the condition for the occurrence of the well known unstable circular null
geodesics by taking the limit $E_{0}\rightarrow \infty$, when
\begin{eqnarray}
Z_{\pm} &=& \left[1-N Mu_{0}^{N-2}+(N-1)Q^2u_{0}^{2N-4}\right] \pm
\nonumber \\[2mm] &&
2a\sqrt{\left[(N-2)Mu_{0}^N-(N-2)Q^2u_{0}^{2N-2}\right]}=0 ~.\label{z0}
\end{eqnarray}
or alternatively for $r_{0}=r_{c}$
\begin{eqnarray}
r_{c}^{2N-4}-N Mr_{c}^{N-2}\pm 2ar_{c}^{N-3}\sqrt{(N-2)Mr_{c}^{N-2}-(N-2)Q^2}+(N-1)Q^2 &=& 0~.\nonumber\\
\label{rul}
\end{eqnarray}
Here $(-)$ sign indicates for direct orbit and $(+)$ sign indicates for retrograde orbit. The real positive root of the
equation is the closest circular photon orbit of the blackhole.

\section{\label{letc} Lyapunov exponent and Timelike Circular Geodesics:}

Now we evaluated the Lyapunov exponent for timelike circular geodesics as follows, using equations (\ref{pot}) we get
\begin{eqnarray}
\lambda_{time} &=& \sqrt{\frac{(N-2)}{r_{0}^{N-2}Z_{\mp0}}[r_{0}^{N-4}(NMr_{0}^{N-2}-2(N-1)Q^2)\Delta_{r_0}-
(4Mr_{0}^{N-2}-Q^2)Z_{\mp0}]} \nonumber \\~.\label{lymp2}
\end{eqnarray}
where
\begin{eqnarray}
\Delta_{r_{0}} &=& r_{0}^{N}+a^2r_{0}^{N-2}-2Mr_{0}^{2}+Q^2r_{0}^{4-N}\nonumber\\
Z_{\mp0} &=& r_{0}^{2N-4}-N Mr_{0}^{N-2}\mp 2ar_{0}^{N-3}\sqrt{(N-2)Mr_{0}^{N-2}-(N-2)Q^2}+(N-1)Q^2. \nonumber
\end{eqnarray}
Since $\Delta_{r_0} \geq Z_{\mp0}\geq 0$ and  for $r_{0}^{N-4}(NMr_{0}^{N-2}-2(N-1)Q^{2})\Delta_{r_0}\geq(4Mr_{0}^{N-2}-4Q^{2}) Z_{\mp0}$, $\lambda$ is real, so we conclude that there are no ISCOs around the charged Myers-Perry blackhole space-time.

\section{\label{cetc} Critical exponent and Timelike Circular Geodesics:}

Now we determine the Critical exponent of charged Myers Perry black hole space-time for equatorial timelike circular geodesics. From that 
we shall prove the instability of equatorial timelike circular geodesics via Critical exponent. Thus the reciprocal of Critical
exponent is given by
\begin{eqnarray}
\frac{1}{\gamma} &=& 2\pi \, (r_{0}^{N-1}\mp a\sqrt{(N-2)Mr_{0}-(N-2)Q^2})\times \nonumber \\[4mm] && \frac{\sqrt{\left[r_{0}^{N-4}(NMr_{0}^{N-2}-2(N-1)Q^{2})\Delta_{r0}
-(4Mr_{0}^{N-2}-4Q^{2})Z_{\mp0}\right]}}{\sqrt{r_{0}^{N-2}(Mr_{0}^{N-2}-Q^2)Z_{\mp0}}} ~.\label{cemp}
\end{eqnarray}
Since $Z_{\mp0}\geq 0$, $\Delta_{r_0}\geq 0$ and $(NMr_{0}^{N-2}-2(N-1)Q^{2})\Delta_{r_0} \geq (4Mr_{0}^{N-2}-4Q^{2})Z_{\mp0}$, so  $\frac{1}{\gamma}$ is real, which also implies that equatorial time like circular geodesics is unstable.

\section{\label{lenc}Lyapunov exponent and Null Circular Geodesics:}

For null circular geodesics the Lyapunov exponent and K-S entropy of charged Myers Perry blackhole are given by
\begin{eqnarray}
\lambda_{null} &=& \sqrt{\frac{(N-2)(L_{c}-aE_{c})^2[NMr_{c}^{N-2}-2(N-1)Q^2]}{r_{c}^{2N}}} ~.\label{lemp2}
\end{eqnarray}

Since $NMr_{c}^{N-2}>2(N-1)Q^2$ therefore $\lambda$ is real so the null circular geodesics are unstable.
In the appropriate limits, we can obtain the Lyapunov exponent
for Myers Perry space-times, Tangherlini RN space-times and Tangherlini Schwarzschild space-times.
\section{\label{cenc}Critical Exponent and Null Circular Geodesics:}

Therefore the reciprocal of Critical exponent is given by
\begin{eqnarray}
\frac{1}{\gamma} &=& 2\pi \, (r_{c}^{N-1}\mp a\sqrt{(N-2)Mr_{c}-(N-2)Q^2}) \times \nonumber \\[4mm] && \sqrt{\frac{(L_{c}-aE_{c})^2[NMr_{c}^{N-2}-2(N-1)Q^2]
}{r_{c}^{2N}(Mr_{c}^{N-2}-Q^2)}} ~.\label{cempn}
\end{eqnarray}

\section{\label{dis} Discussion}

The study demonstrates that Lyapunov exponent  may be used to give a full description of  time-like circular  geodesics
and null circular geodesics in charged Myers Perry black hole space-time. We  showed that the Lyapunov exponent
can be used to determine the instability of equatorial circular geodesics,
both time-like and null case for charged Myers Perry  black hole  space-times. By computing
Lyapunov exponent, we proved that for more than four space time dimensions$(N \geq 3)$, there is no ISCO in
charged Myers Perry black hole spacetimes. The other point we have studied that for circular geodesics around
the central black-hole, time-like circular geodesics is characterized by the smallest angular frequency as measured
by the asymptotic observers-no other circular geodesics can have a smallest angular frequency. Thus such types of
space-times  always have $\Omega_{particle}<\Omega_{photon}$ for all time-like circular geodesics. Alternatively
it was shown that the orbital period for time-like circular geodesics  is  characterized by the longest orbital period
than the null circular geodesics. Hence, we conclude that time-like geodesics provide the \emph{slowest way} to circle
the black hole.

\section*{Acknowledgements}

I would like to thank Prof. P. Majumdar  of R.M.V.U for reading the manuscript.
It is also my great plesure to thank Prof. Kumar Shwetketu Virbhadra  for his comments and
suggestions regarding the ``Photon Sphere''.

\end{document}